\documentclass[conference]{IEEEtran}
\IEEEoverridecommandlockouts
\usepackage{cite}
\usepackage{url}
\usepackage{hyperref}
\usepackage{amsmath,amssymb,amsfonts}
\usepackage{algorithmic}
\usepackage{graphicx}
\usepackage{textcomp}
\usepackage{xcolor}
\def\BibTeX{{\rm B\kern-.05em{\sc i\kern-.025em b}\kern-.08em
    T\kern-.1667em\lower.7ex\hbox{E}\kern-.125emX}}
\begin{document}

\title{PABAU: Privacy Analysis of Biometric API Usage}

\author{\IEEEauthorblockN{Feiyang Tang}
\IEEEauthorblockA{\textit{Norwegian Computing Center} \\
Oslo, Norway \\
feiyang@nr.no}
}

\maketitle

\begin{abstract}
Biometric data privacy is becoming a major concern for many organizations in the age of big data, particularly in the ICT sector, because it may be easily exploited in apps. 
Most apps utilize biometrics by accessing common application programming interfaces (APIs); hence, we aim to categorize their usage. 
The categorization based on behavior may be closely correlated with the sensitive processing of a user's biometric data, hence highlighting crucial biometric data privacy assessment concerns.
We propose PABAU, Privacy Analysis of Biometric API Usage. 
PABAU learns semantic features of methods in biometric APIs and uses them to detect and categorize the usage of biometric API implementation in the software according to their privacy-related behaviors. 
This technique bridges the communication and background knowledge gap between technical and non-technical individuals in organizations by providing an automated method for both parties to acquire a rapid understanding of the essential behaviors of biometric API in apps, as well as future support to data protection officers (DPO) with legal documentation, such as conducting a Data Protection Impact Assessment (DPIA).
\end{abstract}


\section{Introduction}
Authentication solutions are shifting from password-based to passwordless solutions due to the documented risks of passwords because of the user-generated credentials, brute-force attacks, recycled passwords, and large-scale breaches.
Taking advantage of biometrics, biometric authentication solutions are deployed for both professional usage, for example, company intranet authentication, and personal usage, for example, bank app authentication.
The usage diversification of passwordless authentication solutions, based on biometrics, raises privacy concerns since biometric data is acquired and processed within a connected multi-program environment, such as a mobile device or a laptop instead of an offline single-program environment, such as a USB stick~\footnote{\url{https://fidoalliance.org/specifications/}} that uses biometric as the authentication resource.

With the more common usage of biometric authentication, manipulating biometric data becomes controversial and makes those solutions run the risk of exposing this sensitive data.
It is critical to guarantee excellent practice and effective solution implementation throughout the software development process. 
With more and more privacy-focused legal provisions being implemented, such as the General Data Protection Regulation (GDPR) in Europe and the California Consumer Privacy Act (CCPA), many companies are becoming aware of the importance of privacy in software development. 
They are beginning to review their code for potential privacy issues in order to avoid the potential legal risks associated with these newly implemented provisions. 
GDPR classifies biometric data as ``sensitive''~\cite{gdprartical9}, and privacy protection on biometric data has become one of the most discussed topics in privacy-related research~\cite{gruschka2018privacy,jasserand2016legal}. 
Biometric processing is now a significant and vital aspect of privacy protection. 
Given that more software employs biometric APIs to authenticate users for convenience, it is critical to understand how biometric APIs function and how biometric data is processed.
As a result of their nature and sensitive use, biometric APIs must be used with a clear understanding by developers and data protection officers (DPOs) need relevant information from developers.
A rigorous privacy assessment of the biometric authenticators is crucial to mitigate privacy risks.
However, such an assessment is difficult to perform due to the subtle nature of privacy and the complex structure of programs implementing biometric authenticators.

In this paper, we would like to understand how biometric-relevant APIs are implemented in real-world applications. We propose PABAU, a technique to classify the usage of biometric APIs in the applications. 
Various usages (those are, methods from the biometric APIs) are classified into different labels according to their behavior, for example, biometric acquisition, user interaction, data transfer, data erasure, etc. 
We can use the results of the analysis to assist developers in better understanding current solutions that adopt biometric APIs in an action-based analysis and guide them to the section that requires more attention and provide a broad overview of how biometric data is handled and manipulated in the application to either project managers or data protection officers without examining the actual implementation code.

Our contributions are:
\begin{itemize}
    \item An method for automatically categorizing biometric API methods (Section~\ref{sec:algorithm} and Section~\ref{sec:training}).
    \item A handcrafted ground-truth dataset built from the popular Android and FIDO2 biometric authentication APIs (Section~\ref{sec:truth}).
    \item An automatic scheme to describe the privacy behavior of API usage in client code, for instance, authentication, cryptography, termination, and permission, that could aid developers and DPOs in analyzing biometric-related privacy compliance (Section~\ref{sec:labels}).
\end{itemize}

We demonstrate the utility of our research by testing our classifier on eight popular Android applications (Section~\ref{sec:experiment}).

\section{Related work}\label{sec:related}
Our study focuses on utilizing static analysis to discover software vulnerabilities and supporting non-technical experts with software privacy compliance checks. This section begins with a discussion of traditional software security vulnerability discovery using program analysis, followed by a discussion of software privacy assessment research.

Taint analysis is a practical approach to information flow analysis that is frequently used by researchers to analyze the transmission of private data. It comprises both dynamic and static taint analysis techniques. 
Using taint analysis to find privacy vulnerabilities in Android applications has been discussed in multiple research. 

By studying data relationships between program variables without running the program, static taint analysis may determine if data can propagate from a taint source to a taint aggregation point~\cite{wang2017principle}. 
Many Java and Android-based research are based on the tool Soot.
Soot~\cite{vallee2010soot}\footnote{\url{https://github.com/soot-oss/soot}} is a Java bytecode analysis tool developed by the Sable research group of McGill University in 1996. It provides a variety of bytecode analysis and transformation functions, through which it can perform intra- and inter-process analysis and optimization, and program flow analysis.
FlowDroid~\cite{arzt2014flowdroid} mimics a component's lifecycle by building virtual main functions to find sensitive data transfer channels for privacy leaks with an 86\% accuracy.
MudFlow~\cite{avdiienko2015mining} used the static taint analysis tool FlowDroid to test 2,866 benign software and produced a total of 338,610 taint analysis results. The tool can only give out whether a piece of software as a whole has malicious behavior (and the tool will also have false positives) and cannot tell whether a taint analysis path (source to sink) is the result of direct privacy leakage.
Moving from Android to general Java applications, in 2018, Sas et al. proposed SSCM~\cite{sas2018automatic} to identify data sources and sinks from arbitrary Java libraries. 
SSCM uses Soot Framework to extract information from the Java Bytecode and uses its novel rules and WEKA Data Mining Framework~\cite{hall2009weka} to classify different types of security sources and sinks. 
SWAN~\cite{swanPaper}\footnote{\url{https://github.com/secure-software-engineering/swan/tree/master/swan_core}} is similar to SSCM. Both SSCM and SWAN use similar features for the classification model's learning process. SSCM proposed 9 when SWAN contains 25 basic features. These features focus mostly on the name of \textit{Resource Class} and \textit{Resource Method} and \textit{call to the corresponding methods}. Meanwhile, parameters and some specific patterns are also important. Almost all the features used in SSCM have been included in SWAN, and SWAN contains a few more features on return types and some specific patterns in methods. 
Meanwhile, parameters and some specific patterns are also important.

Privacy-by-design (PbD) has inspired a significant amount of research that provides approaches and models to preserve software privacy before implementation begins, as well as to forecast or manage developer privacy compliance throughout implementation~\cite{hadar2018privacy,hoepman2014privacy}. 
According to a survey~\cite{dias2020perceptions} written by Dias Canedo et al., technical employees frequently lack legal expertise in terms of privacy protection, which encouraged us to design a new easy-to-use technique for developers to check the privacy compliance of their implementation on a regular basis without diving into hundreds of lines of code.

With requirements like the data protection impact assessment (DPIA), researchers have been attempting to make it easier for non-technical professionals, such as attorneys, to confirm GDPR compliance. 
To aid the examination of privacy consequences in ubiquitous computing systems, Fern{\'a}ndez et al.~\cite{fernandez2019software} suggested a software-assisted process methodology. Similarly, the BPR4GDPR~\cite{lioudakis2019facilitating} initiative has the same objective, while Zibuschka~\cite{zibuschka2019analysis} analyzed the automation potential of the DPIA process. 
Using privacy flow-graphs, Tang et al.~\cite{tang2022assessing} presented an approach for analyzing privacy in the context of GDPR.
This motivates us to bridge the gap between the technical and non-technical experts with an approach that might assist both parties in terms of assessing privacy in biometric API usage.

\section{Motivation} \label{sec:motiv}
Conducting a privacy analysis necessitates a vast number of software information, many of which might be highly specific.
Privacy lawyers may request that developers offer precise answers with particular processing in software, which could be challenging when the software is not built and maintained by a single team from the start.
Furthermore, manually reviewing hundreds of lines of code demands a significant amount of effort on the part of the development team.
It is difficult for developers and lawyers to have a mutual understanding and benefit from each other's expertise and analyzes~\cite{hansen2011top,hadar2018privacy,bettini2015privacy}.
Our goal is to see if we can transpose GDPR principles and requirements with actual code locations or patterns in software.
In this section, we look at fundamental GDPR obligations from the standpoint of a developer and then consider how we may change our solution to serve both parties in terms of privacy protection in biometric API usage.

\subsection{Legal perspectives}\label{sec:legal}

GDPR states several obligations from the data controller which should be monitored by the DPO~\cite{gdpr}:
\begin{itemize}
    \item by default and by design, to have a record of processing activities (Article 30);
    \item to ensure the security of the processing (Article 32);
    \item to notify personal data breaches to the supervisory authorities (Article 33);
    \item to communicate personal breaches to the data subject (Article 34);
    \item to conduct DPIA (Article 35);
    \item to conduct prior consultation with supervisory authorities (Article 36).
\end{itemize}

The DPO's mission is to monitor whether the data controller fulfilled all of his or her commitments, which includes performing a high-quality DPIA when required. This makes the assignment of DPIA creation a duty for both data controllers and DPOs.
DPIA is a technique that assists developers and organizations in systematically analyzing, identifying, and mitigating data protection risks in a project or plan. 
The deliverable document created by the approach in this study is referred to as a DPIA. 

We intend to offer tailored solutions for DPIA under diverse applications, similar to earlier research~\cite{horak2019gdpr,henriksen2020dpia}.
The aim of our technique is to provide information that can directly benefit developers and DPOs by providing comprehensive guidance on producing a quality DPIA. 
Developers and DPOs may identify the components of the software that process biometric data from users within applicable legal viewpoints with a list of biometric API use described under different kinds of behavioral labels. Because various API methods can capture different forms of biometrics processing, we can adopt this into our study.

\section{Approach}\label{sec:method}
To tackle the barrier between developers and DPOs in terms of biometric privacy protection compliance checks, we utilize SWAN as a foundation to create a technique PABAU that can automatically categorize methods in biometric APIs based on their actions.

\subsection{General architecture}
\label{sec:algorithm}
\begin{figure*}[h!]
  \centering
  \includegraphics[width=\textwidth]{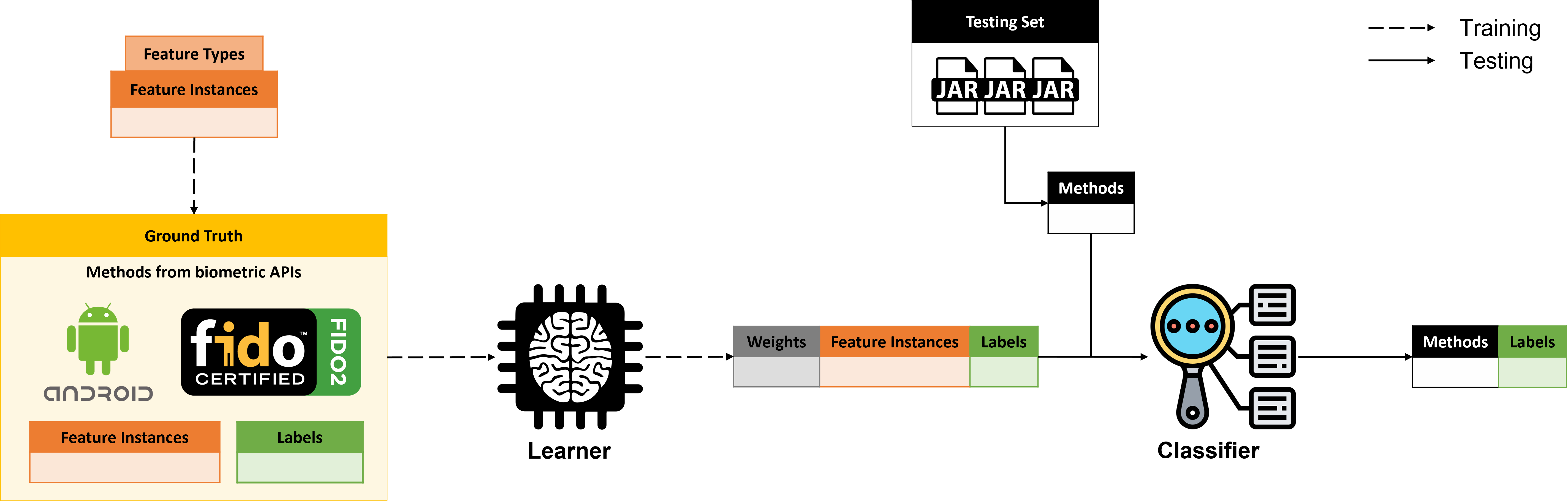}
  \caption{General architecture of PABAU}
  \label{fig:Overview}
\end{figure*}

The architecture of our technique is illustrated in Fig.~\ref{fig:Overview}. 
The compiled class files from Java or Android applications are our source of analysis. 
We can learn the features of these APIs and use them to categorize their usage in real-world applications by training a model based on the most common biometric API methods, such as the Android official biometric API and FIDO2 implementations.
The biometric types included in this study cover the two most common biometrics: face and fingerprint.

PABAU operates on different classification sets, one for each kind of biometric API method. 
Most action-based sets are not disjoint (for example, \texttt{Authenticate()} may be labeled as both \textbf{CRYPTO} (cryptography involved) and \textbf{AUTHENTICATE}), but certain sets like the biometric strength levels (for example, \textbf{BSC1} and \textbf{BSC2}) are disjoint (that is, each method can only get labeled as one of them). 
The classification for each label operates independently.

\subsection{Feature types and feature instances}\label{sec:feature}
We adopted the same system as SWAN with feature types and corresponding feature instances. The features are extracted from the semantics of methods in the program. Feature types are general features such as \texttt{methodNameContains} and feature instances represent concrete instances of feature types such as \texttt{methodNameContainsUser}. We considered the following criteria when constructing feature types.

\begin{itemize}
    \item Names of the methods and the declaring classes: \begin{itemize}
        \item Start with a keyword;
        \item Contain a keyword;
        \item End with a keyword.
    \end{itemize}
    \item Return types of the methods: \begin{itemize}
        \item Primitive types (for example, \texttt{int}, \texttt{string});
        \item Other objects (for example, \texttt{PromptInfo}).
    \end{itemize}
    \item Parameters: \begin{itemize}
        \item Number of parameters;
        \item Types of parameters: primitive or other object types.
    \end{itemize}
    \item Invocations: whether the current method calls any other method: \begin{itemize}
        \item The name of callee;
        \item The return type of callee;
        \item The parameter of the callee.
    \end{itemize}
    \item Particular pattern derived from above: \begin{itemize}
        \item Parameter flows into the return statement;
        \item Parameter flows into a local field;
        \item A local field flows into the return statement.
    \end{itemize}
\end{itemize}

\subsection{Labels}\label{sec:labels}
Some particular actions of biometric APIs might lead to potential privacy risks, which motivates us to create labels representing different actions.
The following labels in Tab.~\ref{tab:label} are what we used to classify different behaviors in biometric API usage.

\begin{table}[h!]
\centering
\caption{Labels and the description}
\label{tab:label}
\resizebox{\columnwidth}{!}{%
\begin{tabular}{ll}
\multicolumn{1}{c}{\textbf{Label}} & \multicolumn{1}{c}{\textbf{Description}}                                  \\ \hline\hline
BSC1        & \begin{tabular}[c]{@{}l@{}}Biometric strength level `convenience'. \\ This level does not use cryptography nor the \texttt{BiometricPrompt} API.\end{tabular} \\ \hline
BSC2        & \begin{tabular}[c]{@{}l@{}}Biometric strength level `weak'. \\ This level uses only the \texttt{BiometricPrompt} API but no cryptography.\end{tabular}        \\ \hline
BSC3        & \begin{tabular}[c]{@{}l@{}}Biometric strength level `strong'. \\ This level uses both the\texttt{BiometricPrompt} API and cryptography.\end{tabular}         \\ \hline
SOURCE                              & Where the potentially sensitive data come from.                            \\ \hline
SINK                                & The places where the sensitive data might end up in.                       \\ \hline
CHECKER                             & Checking occurs, for example, check for hardware prerequisites.            \\ \hline
PERMISSION                          & Getting or verifying the user's permission for the biometric process.      \\ \hline
AUTHENTICATE                        & Authentication-related process.                                            \\ \hline
CRYPTO                              & Where encryption is involved in the process.                               \\ \hline
TERMINATION                         & Where termination of specific service(s) happens.                          \\ \hline
INTERACTION & \begin{tabular}[c]{@{}l@{}}The system is making interactions with users, \\ for example, giving users a few options.\end{tabular}                         \\ \hline
TRANSFER                            & Biometric-related data or decisions derived from it are being transferred. \\ \hline
ACQUISITION                         & The acquisition of biometric-related data.                                 \\ \hline
DELETION                            & The deletion of biometric-related data.                                    \\ \hline
STORAGE                             & Where biometric-related data is being stored and kept.                     \\ \hline
DATABASE                            & Where a database is involved.                                              \\ \hline
\end{tabular}%
}
\end{table}

\subsection{Ground-truth}\label{subsubsection:groundTruth}
\label{sec:truth}
In order to train PABAU, we need to build a ground-truth dataset specific to biometric authentication and annotate the methods. 
Because Android apps are the most popular domain for biometric API usage, we chose official Android Biometric APIs as our domain of study. 
To also cover web applications that use FIDO2, we also find a representative implementation to enrich the ground truth.

Initially, we parse the existing methods from Android biometrics APIs\footnote{\url{https://android.googlesource.com/platform/frameworks}} and the FIDO2 server implementation from LINE\footnote{\url{https://github.com/line/line-fido2-server}}.
To enrich the training set, we also manually annotated over 150 methods from several popular sample implementations of biometric APIs\footnote{Soter: \url{https://github.com/Tencent/soter/}}\footnote{Android-Goldfinger
: \url{https://github.com/infinum/Android-Goldfinger}}\footnote{\url{https://github.com/sergeykomlach/AdvancedBiometricPromptCompat}}. 
We manually decompress the library files, which were originally in JAR format, and label the methods in the class files in accordance with their corresponding behaviors. 
A general training set should cover all of the common biometric API methods, which is why we include the native Android biometric API methods and some third-party ones. Each of these methods is annotated with privacy behavior labels.
Then, in a structured manner, for each method we collect the following properties: 
\begin{itemize}
    \item \textbf{name.} The method's fully qualified name (for example, \texttt{package.class.method}). This indicates which class the method belongs to.
    \item \textbf{return.} The method's return type (\texttt{void}, primitive types (for example, \texttt{int}), or objects of other classes). This helps determine whether there is biometric data flow out from a local method.
    \item \textbf{parametersTypes.} The method's parameters types (empty, primitive types, for example, \texttt{int}), or reference types).
    \item \textbf{calleeNames.} The full name of the methods invoked inside that method. This helps determine relationships between methods, to locate data flows of biometric data better.
\end{itemize}

\begin{figure}[h!]
  \centering
  \includegraphics[width=\columnwidth]{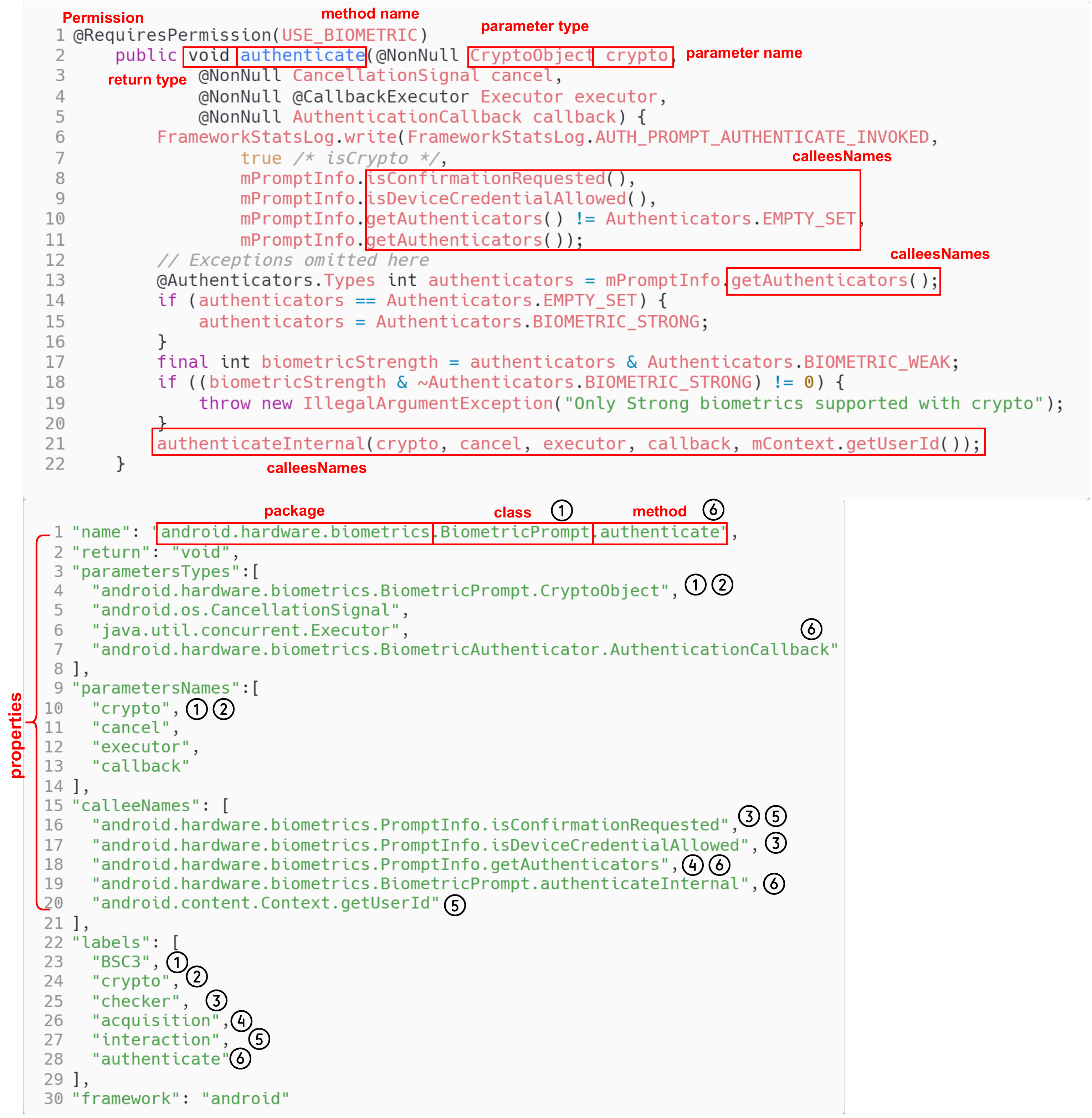}
  \caption{Example of a method from the Android Biometric API: \texttt{authenticate} is a method from the class \texttt{android.hardware.biometrics.BiometricPrompt} and the corresponding annotated method (in the green text) data point. The numbered circles (from lines 23 to 28) correspond to the labels (Section~\ref{sec:labels}). They were assigned to this data point since \texttt{authenticate} has the properties in lines 1, 4, 7, 10, 16-20 in the text.}
  \label{fig:authenticate-java}
\end{figure}

\begin{figure*}[!h]
  \centering
  \includegraphics[width=.9\textwidth]{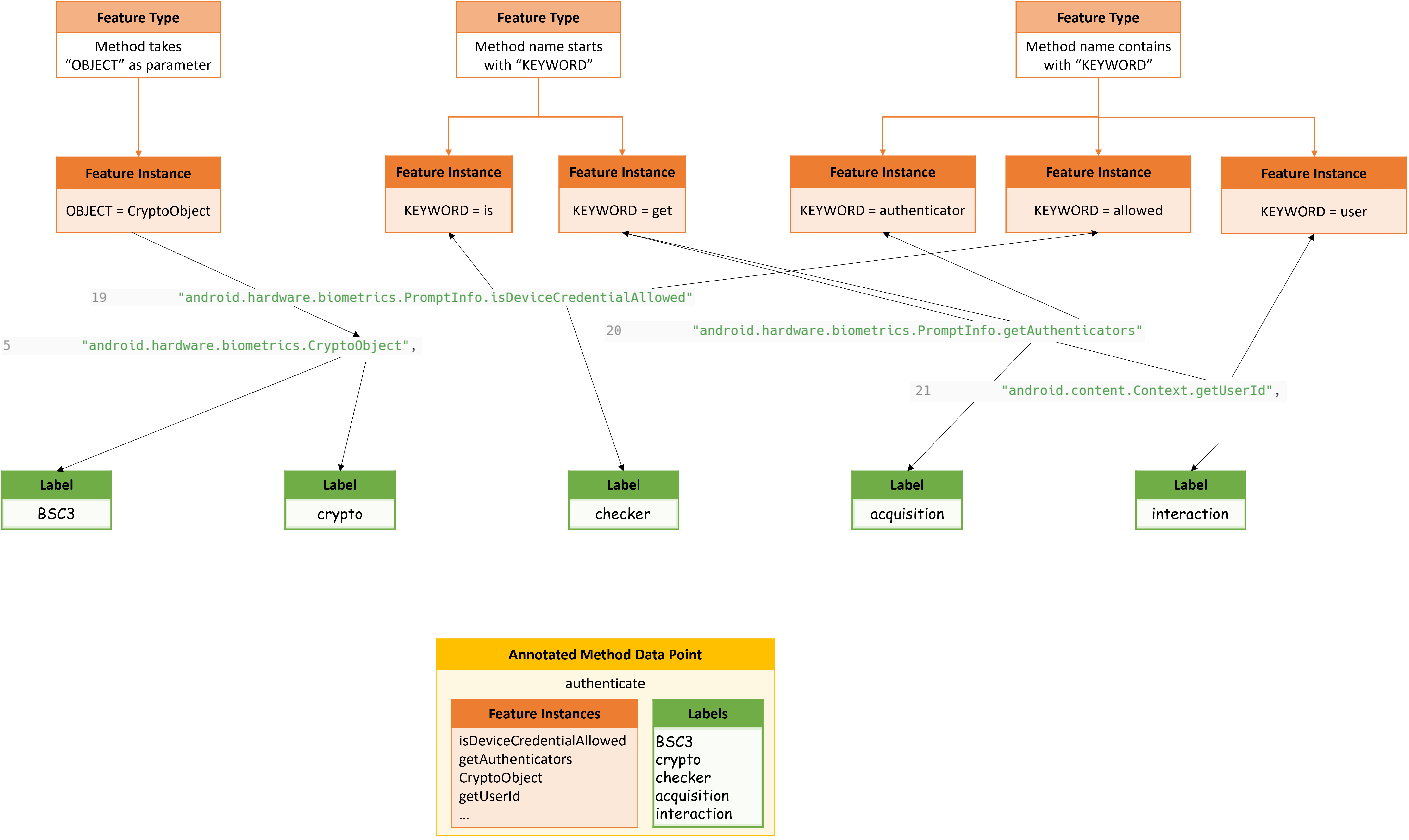}
  \caption{Annotation diagram of the sample method}
  \label{fig:annotation}
\end{figure*}

Each method in this dataset is manually associated with the label(s), the detailed explanation of labels is discussed in Section~\ref{sec:labels}. 
The example of how the training process is illustrated in Fig.~\ref{fig:authenticate-java} along with an annotation example in Fig.~\ref{fig:annotation}. Details on which type of feature and feature instance are summarized in Section~\ref{sec:feature}.

\section{Training and classification}\label{sec:training}
Similarly to SWAN~\cite{swanPaper}'s technique, we develop a collection of binary features that assess certain qualities of the methods to assist the machine learning algorithm in labeling the methods based on their actions. 

We evaluated six popular classifiers from WEKA~\cite{hall2009weka}, a commonly used machine learning API written in Java: Bayes Net, Naive Bayes, Logistic Regression, C4.5, and Decision Stump and SVM.
For each classifier in the training set, ten 10-fold cross-validations were conducted to find the best classifier for PABAU.
The median precision and recall statistics for each classifier are presented in Table.~\ref{tab:PRdiffalg}.
In terms of average accuracy and recall, we can observe that SVM performed the best with an average precision of 0.9725 and recall of 0.98. 
Except for Stump, which performed far less well than the other five, the majority of classifiers functioned well.

For example, the feature instance \texttt{hasCalleeNameStartsCheck} is likely to indicate that the current method calls a checker function to see whether specific conditions have been satisfied. 
When the learning process starts, PABAU computes a yes or no response for each feature instance for each method in the training set. 
Based on the feedback, we are able to learn which combination of features is the most associated with the labels. 
These combinations will be then used to label methods in testing sets.

\begin{table*}[ht]
\centering
\caption{Precision (P) and recall (R) of the 10-cross fold validation for all classifiers averaged over 10 iterations}
\label{tab:PRdiffalg}
\begin{tabular}{l|cccccccccc}
 &
  \multicolumn{2}{c}{Source} &
  \multicolumn{2}{c}{Sink} &
  \multicolumn{2}{c}{Auth} &
  \multicolumn{2}{c}{Crypto} &
  \multicolumn{2}{c}{Average} \\
 &
  P &
  \multicolumn{1}{c|}{R} &
  P &
  \multicolumn{1}{c|}{R} &
  P &
  \multicolumn{1}{c|}{R} &
  P &
  \multicolumn{1}{c|}{R} &
  P &
  R \\ \hline\hline
BayesNet &
  0.92 &
  \multicolumn{1}{c|}{0.97} &
  0.95 &
  \multicolumn{1}{c|}{0.95} &
  1.00 &
  \multicolumn{1}{c|}{1.00} &
  1.00 &
  \multicolumn{1}{c|}{1.00} &
  0.9675 &
  0.9800 \\
NaiveBayes &
  0.93 &
  \multicolumn{1}{c|}{0.95} &
  0.94 &
  \multicolumn{1}{c|}{0.96} &
  0.88 &
  \multicolumn{1}{c|}{1.00} &
  1.00 &
  \multicolumn{1}{c|}{1.00} &
  0.9375 &
  0.9775 \\
Logistic Reg &
  0.94 &
  \multicolumn{1}{c|}{0.95} &
  0.95 &
  \multicolumn{1}{c|}{0.92} &
  0.88 &
  \multicolumn{1}{c|}{1.00} &
  1.00 &
  \multicolumn{1}{c|}{1.00} &
  0.9425 &
  0.9675 \\
C4.5 &
  0.93 &
  \multicolumn{1}{c|}{0.96} &
  0.93 &
  \multicolumn{1}{c|}{0.95} &
  0.88 &
  \multicolumn{1}{c|}{1.00} &
  1.00 &
  \multicolumn{1}{c|}{0.75} &
  0.9350 &
  0.9775 \\
Stump &
  0.79 &
  \multicolumn{1}{c|}{0.84} &
  0.88 &
  \multicolumn{1}{c|}{0.92} &
  0.75 &
  \multicolumn{1}{c|}{1.00} &
  1.00 &
  \multicolumn{1}{c|}{0.5} &
  0.8550 &
  0.8150 \\
SVM &
  0.94 &
  \multicolumn{1}{c|}{0.96} &
  0.95 &
  \multicolumn{1}{c|}{0.96} &
  1.00 &
  \multicolumn{1}{c|}{1.00} &
  1.00 &
  \multicolumn{1}{c|}{1.00} &
  \textbf{0.9725} &
  \textbf{0.9800}
\end{tabular}%
\end{table*}

\section{Experiment}\label{sec:experiment}
The main research question here is how our technique performs in terms of assigning biometric API methods with the corresponding behavioral labels. 
We aim for high precision because calculating recall is unrealistically tricky. 
All our experiments were performed on a Windows machine with an Intel i7 1.90GHz CPU and 16 GB memory.

\subsection{Description of the datasets}
Besides the ground-truth dataset we built using Android biometric APIs and FIDO2 server implementation, we also manually annotated biometric API usages from real-world Android mobile applications.
Given that biometric authentication is only used in a subset of Android applications, we chose eight popular apps from the Google Play Store\footnote{\url{https://play.google.com/store/apps/}} in four categories: banking \& finance (Klarna, Sparebank1, Revolut, Paypal), private storage (Private Photo Vault, NordLocker), privacy notebook (Notability), and password management (1Password).

The description of the main functionality and biometric usage of the apps are listed in Table~\ref{tab:datasets}.

We split the entire handcrafted dataset into two parts, 70\% of the data points were used to train the classifier and the rest 30\% were used for testing.

\begin{table}[h!]
\centering
\caption{Description of the datasets}
\label{tab:datasets}
\resizebox{\columnwidth}{!}{%
\begin{tabular}{ll}
\multicolumn{1}{c}{\textbf{App name}} &
  \multicolumn{1}{c}{\textbf{Description}} \\ \hline\hline
Klarna &
  \begin{tabular}[c]{@{}l@{}}A financial app that provides online payment and \\ direct payments along with post-purchase payments.\\ Users can use biometrics to unlock the app \\ and authorise certain payments.\end{tabular} \\ \hline
Sparebank1 &
  \begin{tabular}[c]{@{}l@{}}A online banking app for Sparebank1. \\ Users can use biometrics to unlock the app, \\ authorize certain payments and approve e-invoice.\end{tabular} \\\hline
Revolut &
  \begin{tabular}[c]{@{}l@{}}An online banking app for Revolut Bank.\\ Users can use biometrics to unlock the app, add a \\ new card, authorize payments, and update profiles.\end{tabular} \\\hline
Paypal &
  \begin{tabular}[c]{@{}l@{}}A financial app that provides the online payment \\ between users or merchants.\\ Users can use biometrics to unlock the app, \\ authorize payments, and update card/account details.\end{tabular} \\\hline
Private Photo Vault &
  \begin{tabular}[c]{@{}l@{}}A photo storage app that provides access control.\\ Users can use biometrics to unlock the app, add photos \\ to the vault and add extra control to certain folders.\end{tabular} \\\hline
NordLocker &
  \begin{tabular}[c]{@{}l@{}}A storage app that provides access control.\\ Users can use biometrics to unlock the app, add files into \\ the locker and apply extra encryption to certain files.\end{tabular} \\\hline
Notability &
  \begin{tabular}[c]{@{}l@{}}An online notebook.\\ Users can use biometrics to lock and unlock certain notes.\end{tabular} \\\hline
1Password &
  \begin{tabular}[c]{@{}l@{}}A password manager.\\ Users can use biometrics to unlock the app, add/update \\ credentials, and sync information with the cloud.\end{tabular} \\\hline
\end{tabular}%
}
\end{table}

\subsection{Example}
\label{sec:eg}
\begin{figure*}[h!]
  \centering
  \includegraphics[width=\textwidth]{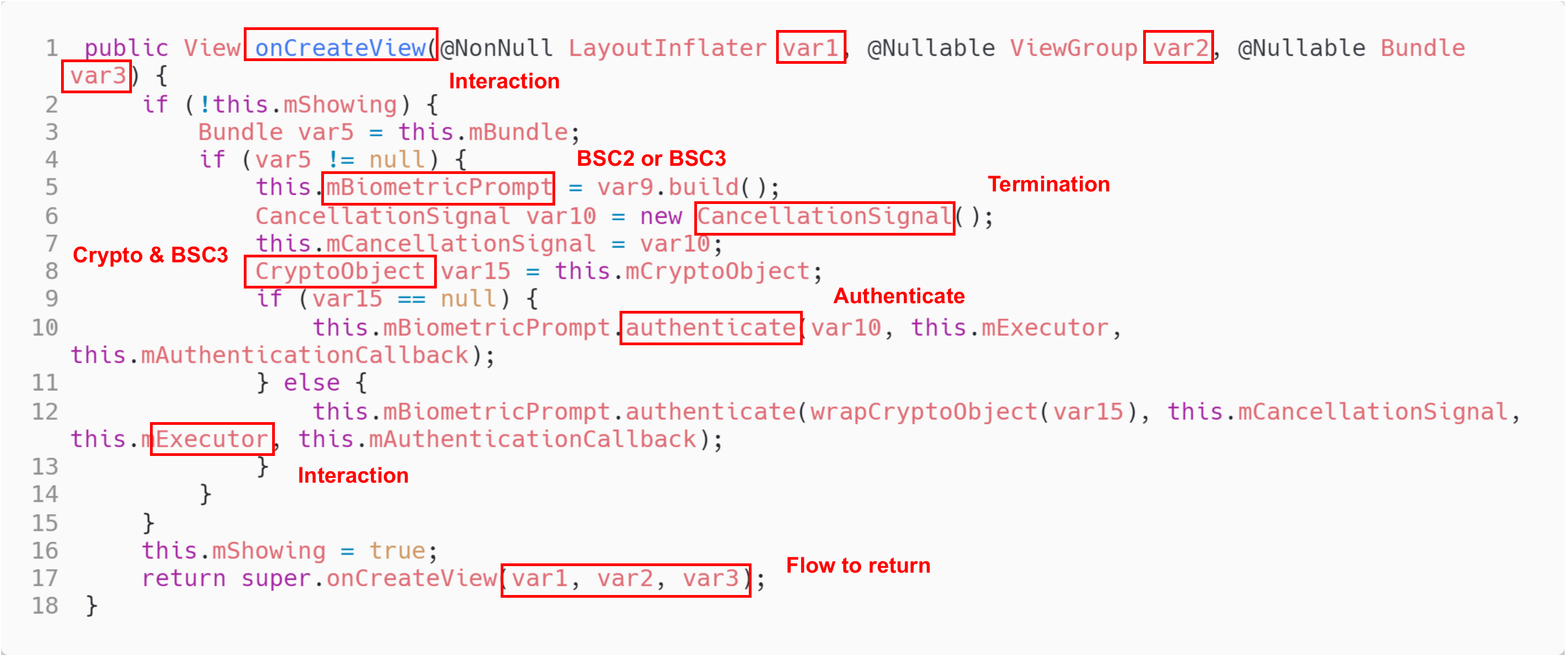}
  \caption{Sample classification result snippet from Sparebank1}
  \label{fig:sb1eg}
\end{figure*}

Fig.~\ref{fig:sb1eg} gives an example of what kind of result PABAU generates by analyzing the JAR files.
The method \texttt{onCreateView} was extracted from the class \\\texttt{androidx.biometrics.BiometricFragment}.

We could see that the method itself gets classified by multiple labels, for example, it gets classified as \textbf{BSC3} and \textbf{Crypto} from the invocation to the data type \texttt{CryptoObject}.
The app's implementation has a fairly small number of biometric API methods, however, this does not diminish their importance.
Actually, their significance might be underestimated, and the underlying correlations between one method and other subtle behaviors reveal a sensitivity to privacy.
This is also why we need a technique such as PABAU to classify each method rigorously with as many privacy behavior labels as feasible.

\subsection{Runtime and memory performance}
Table~\ref{tab:tm} demonstrates the stability of PABAU, and it does not require huge computational costs to analyze common mobile apps. 
The cost of time and memory increases as the size of a program grows, which also reflects a greater number of methods inside the application.
We also observed that the quantity of biometric-related approaches influences the time and memory requirements.
Even though Notability is almost three times the size of Sparebank1, it does not cost three times as much in terms of time and memory since Notability uses biometric APIs for fewer features than Sparebank1.

\begin{table}[h!]
\centering
\caption{Average runtime and memory usage for classifying different apps}
\label{tab:tm}
\resizebox{\columnwidth}{!}{%
\begin{tabular}{l|c|cc}
                    & \#Total       & \multicolumn{2}{c}{Average cost} \\
                    & methods        & Time (s)       & Memory (MB)     \\ \hline\hline
Klarna              & 105,185 & 38.51 $\pm$ 1.68 & 641.48 $\pm$ 2.15 \\
Sparebank1          & 89,142  & 32.79 $\pm$ 0.72 & 584.37 $\pm$ 1.93 \\
Revolut             & 86,457  & 30.25 $\pm$ 0.78 & 518.56 $\pm$ 1.84 \\
Paypal              & 94,618  & 35.96 $\pm$ 1.23 & 604.62 $\pm$ 1.89 \\
Private Photo Vault & 175,841 & 43.19 $\pm$ 1.48 & 659.07 $\pm$ 2.24 \\
NordLocker          & 115,294 & 30.57 $\pm$ 1.55 & 556.29 $\pm$ 2.05 \\
Notability          & 284,585 & 65.48 $\pm$ 2.14 & 844.73 $\pm$ 3.21 \\
1Password           & 134,806 & 31.33 $\pm$ 1.67 & 581.46 $\pm$ 2.23
\end{tabular}%
}
\end{table}

\subsection{Evaluating the precision}
To examine the classification precision, we manually go through the methods that got classified by each behavioral label. 
The methods were extracted from the decompiled \texttt{apk} files and read by our technique.
Table~\ref{tab:result} displays the total number of methods analyzed in each application as well as the classification results for various behaviors (categories). 
It is worth noting that all finance apps use the cryptography-related API, and none of the apps have access to actual biometric data storage, database, or acquisition (the only acquisition that happens is the cryptographic keys involved in the authentication process, no biometric data involved).
This is aligned with GDPR requirements since transporting and storing biometric data is very sensitive and requires further consent from data owners.

According to Table~\ref{tab:result2}, PABAU has an average precision of 0.84 across all the labels. 
It is more precise for detecting \textbf{Source} and \textbf{Sink} (0.96), as well as some categories such as \textbf{BSC1/2/3} (0.97), than for other categories such as \textbf{Termination} (0.87). 
Some of the behavior labels appear infrequently in the dataset as well.

PABAU can be improved by using a larger and more diverse training set that includes real-world API usage annotations and domain-specific information.

\begin{table}[ht]
\centering
\caption{Total number of methods analyzed (\#M) and numbers of methods detected by PABAU per behavioral label. }
\label{tab:result}
\resizebox{\columnwidth}{!}{%
\begin{tabular}{lcccccccc}
           & Klarna  & Sparebank1 & Revolut & Paypal & PPV                 & Nordlocker & Notability & 1Password \\ \hline\hline
\#M        & 105,185 & 89,142     & 86,457  & 94,618 & 175,841             & 115,294    & 284,585    & 134,806   \\
Source     & 127     & 94         & 174     & 140    & 65                  & 98         & 39         & 55        \\
Sink       & 96      & 70         & 143     & 109    & 41                  & 64         & 17         & 32        \\
BSC1       & 15      & 12         & 21      & 16     & 15                  & 20         & 11         & 13        \\
BSC2       & 17      & 24         & 29      & 14     & 7                   & 9          & 0          & 5         \\
BSC3       & 2       & 2          & 3       & 2      & 0                   & 0          & 0          & 0         \\
Checker    & 9       & 7          & 13      & 8      & 3                   & 4          & 1          & 3         \\
Permission & 7       & 5          & 9       & 8      & 1                   & 1          & 1          & 3         \\
Auth       & 8       & 6          & 9       & 11     & 2                   & 3          & 2          & 4         \\
Crypto     & 5       & 5          & 6       & 9      & 0                   & 0          & 0          & 0         \\
Termin     & 9       & 7          & 10      & 9      & 3                   & 1          & 1          & 5         \\
Interact    & 4       & 2          & 5       & 7      & 1                   & 2          & 1          & 4         \\
Transfer   & 0       & 0          & 0       & 0      & 0                   & 0          & 0          & 0         \\
Acquist    & 1       & 1          & 1       & 1      & 0                   & 0          & 0          & 0         \\
Delete     & 2       & 1          & 2       & 3      & 0                   & 0          & 0          & 0         \\
Storage    & 0       & 0          & 0       & 0      & 0                   & 0          & 0          & 0        
\end{tabular}
}
\end{table}

\begin{table}[ht]
\centering
\caption{Number of biometric-related methods detected (\#BM), and precision of PABAU for each label on eight Android applications. `/' marks labels for which PABAU detected no methods.}
\label{tab:result2}
\resizebox{\columnwidth}{!}{%
\begin{tabular}{lcccccccc}
           & Klarna & Sparebank1 & Revolut & Paypal & PPV  & Nordlocker & Notability & 1Password \\ \hline\hline
\#BM       & 326    & 201        & 427     & 372    & 170  & 233        & 89         & 143       \\
Source     & 0.95   & 0.94       & 0.92    & 0.9    & 0.99 & 0.99       & 0.97       & 0.99      \\
Sink       & 0.97   & 0.95       & 0.97    & 0.94   & 0.98 & 0.99       & 0.99       & 0.96      \\
BSC1       & 0.93   & 1          & 1       & 0.89   & 1    & 0.95       & 1          & 0.92      \\
BSC2       & 1      & 0.96       & 0.93    & 1      & 1    & 0.89       & /          & 1         \\
BSC3       & 1      & 1          & 1       & 1      & /    & /          & /          & /         \\
Checker    & 0.67   & 0.57       & 0.77    & 0.5    & 0.33 & 0.5        & 1          & 0.67      \\
Permission & 1      & 0.8        & 0.67    & 0.89   & 1    & 1          & 1          & 1         \\
Auth       & 1      & 1          & 1       & 1      & 1    & 1          & 1          & 1         \\
Crypto     & 1      & 1          & 1       & 1      & /    & /          & /          & /         \\
Termin     & 0.56   & 0.86       & 0.8     & 1      & 1    & 1          & 1          & 0.8       \\
Interact    & 1      & 1          & 1       & 0.86   & 1    & 1          & 1          & 1         \\
Transfer   & /      & /          & /       & /      & /    & /          & /          & /         \\
Acquist    & 1      & 1          & 1       & 0      & /    & /          & /          & /         \\
Delete     & 0.5    & 1          & 1       & 0.3    & /    & /          & /          & /         \\
Storage    & /      & /          & /       & /      & /    & /          & /          & /        
\end{tabular}%
}
\end{table}

\section{Threats to validity}
To check for GDPR compliance in the biometric authentication system and find relevant behavior, here we picked one of the most often used sample templates for generating a DPIA from the Commission nationale de l'informatique et des libertés (CNIL)~\cite{PrivacyI68:online}.

Under ``\textit{Section 2 Assessment of controls protecting data subjects' rights}'', there are key questions that can be answered by the classification results from PABAU. We pick the following examples which can be answered by our classification result:
\begin{itemize}
    \item \textit{Determination and description of the controls for obtaining consent.}

    Typically, the operating system, such as Apple or Google, obtains biometric permission first from users when they enroll their  biometrics into the system. Permission is required the first time an application asks for access to the biometric API. This refers to the methods we classified with the \textbf{Permission} label, these methods indicate the code location and description of biometric permissions sought by the application. From the preceding experiment result, it is evident that all apps that implemented biometric authentication requested permission. By analyzing the discovered methods, further information can be uncovered. For instance, the use of \textbf{Permission} methods may be used to examine privacy policies to determine whether there are locations that utilize biometric data without user consent.
    \item \textit{Information on the secure data storage method, particularly in the event of sourcing.}

    Various biometric security class levels signify varying degrees of security. DPOs need to understand how user data is being securely stored. Labels \textbf{Crypto} and \textbf{BS3} offer the greatest degree of data protection and inform DPOs that the biometric authentication method utilized sophisticated encryption.
    \item \textit{Detailed presentation of the data processing purposes (specified objectives, data matching where applicable, etc.).}

    Biometric APIs are utilized for more than just app unlocking. The label \textbf{Iteract} classifies the interaction between the user and the application in terms of biometric API usage, as user interaction must be precisely identified. By analyzing these approaches, DPOs can determine which activity motivates the use of biometric APIs. Similarly, labels such as \textbf{Auth} explain the authentication procedure in a clear manner by pinpointing the locations of authentication processes and associating the context in the code.
    \item \textit{Possibility of retrieving, in an easily reusable format, personal data provided by the user, so as to transfer them to another service}

    Transferring biometric data or the authentication session's decisions is a highly sensitive process. In spite of the fact that our experiment did not reveal any methods with the \textbf{Transfer} label, it is still worthwhile to discover them in the implementations. Considering the Android biometric data is stored in the Trusty TEE (Trusted Execution Environment)~\cite{enwiki:1105563332}, which is isolated from the rest of the system by both hardware and software. By analyzing the Android implementation, we should not find any data transportation process since Trusty and Android run parallel to each other.
    \item \textit{Indication of the personal data that will nevertheless be stored (technical requirements, legal obligations, etc.).}

    The \textbf{Storage} and \textbf{Delete} labels enable DPOs to identify and comprehend methods that may breach GDPR data retention requirements.
\end{itemize}

\section{Conclusion and future work}\label{sec:conclude}
It has always been a barrier between technical and non-technical people in terms of privacy protection for biometric data. 
DPOs rely on technical specifications from developers to make legal decisions in order to continue monitoring privacy protection compliance in software implementations. 

In this paper, we propose a technique for labeling biometric API usage with behavioral labels. 
This approach overcomes this barrier by providing an automated method for both parties to gain a quick overview of the important fundamental behaviors of biometric API in applications, as well as future assistance to DPOs with legal paperwork, such as performing a DPIA.

Our work provides an early promising result in categorizing the behaviors of biometric API methods in eight popular apps. 
However, we did not have a large number of samples for the training set, the diversity can also be improved. 
Meanwhile, experienced developers' involvement might benefit the learning process by providing valuable relevant elements for training, such as varied weights for features. 
We hope that this method will help both technical and non-technical workers gain a better grasp of privacy protection scenarios during the software development process.

\section*{Acknowledgment}
We appreciate the insightful remarks made by Bjarte M. \O stvold and Amina Bassit. This work is part of the Privacy Matters (PriMa) project. The PriMa project has received funding from European Union’s Horizon 2020 research and innovation program under the Marie Skłodowska-Curie grant agreement No. 860315.
\bibliographystyle{IEEEtran}
\bibliography{ref}

\end{document}